\def\year{2022}\relax
\documentclass[letterpaper]{article} % DO NOT CHANGE THIS
\usepackage{aaai22}  % DO NOT CHANGE THIS
\usepackage{times}  % DO NOT CHANGE THIS
\usepackage{helvet}  % DO NOT CHANGE THIS
\usepackage{courier}  % DO NOT CHANGE THIS
\usepackage[hyphens]{url}  % DO NOT CHANGE THIS
\usepackage{graphicx} % DO NOT CHANGE THIS
\usepackage{natbib}  % DO NOT CHANGE THIS AND DO NOT ADD ANY OPTIONS TO IT
\usepackage{caption} % DO NOT CHANGE THIS AND DO NOT ADD ANY OPTIONS TO IT
\DeclareCaptionStyle{ruled}{labelfont=normalfont,labelsep=colon,strut=off} % DO NOT CHANGE THIS
\usepackage{algorithm}
\usepackage{algorithmic}
\usepackage{amsfonts,bm}
\usepackage{xspace}
\def\ie{{\em i.e.,}\xspace}

\def\etc{{\em etc.,}\xspace}

\def\figref#1{Figure~\ref{#1}}
\def\secref#1{Section~\ref{#1}}
\newcommand{\tabref}[1]{Table~\ref{#1}\xspace}
\newcommand{\equref}[1]{Equation~\ref{#1}\xspace}

\def\1{\bm{1}}

\def\gS{{\mathcal{S}}}

\def\gV{{\mathcal{V}}}

\usepackage{newfloat}
\usepackage{listings}
\floatstyle{ruled}
\newfloat{listing}{tb}{lst}{}
\floatname{listing}{Listing}
\usepackage{microtype}
\usepackage{array}
\usepackage{booktabs}
\usepackage{amsmath}
\usepackage{todonotes}
\usepackage{mathtools}

\usepackage{color} %(if used in text)

\title{Protecting Intellectual Property of Language Generation APIs \\ with Lexical Watermark}
\author {
    Xuanli He$^\dagger$, Qiongkai Xu$^\ddagger$, Lingjuan Lyu$^\mathsection$, Fangzhao Wu$^\sharp$, Chenguang Wang$^\diamond$

}
\affiliations {
    $^\dagger$ Monash University, xuanli.he1@monash.edu\\
    $^\ddagger$ The Australian National University, Qiongkai.Xu@anu.edu.au\footnote{Corresponding authors}\\
$^\mathsection$ Sony AI, Lingjuan.Lv@sony.com 
\footnotemark[1]\\
$^\sharp$ Microsoft Research Asia, wufangzhao@gmail.com\\
$^\diamond$ UC Berkeley, wangcg.pku@gmail.com

}

\usepackage{bibentry}
\begin{document}
% \linenumbers
\maketitle

\begin{abstract}
Nowadays, due to the breakthrough in natural language generation (NLG), including machine translation, document summarization, image captioning, \etc NLG models have been encapsulated in cloud APIs to serve over half a billion people worldwide and process over one hundred billion word generations per day\footnote{\url{https://www.scientific-editing.info/blog/everything-you-need-to-know-about-google-translate/}}. Thus, NLG APIs have already become essential profitable services in many commercial companies. Due to the substantial financial and intellectual investments, service providers adopt a pay-as-you-use policy to promote sustainable market growth. However, recent works have shown that cloud platforms suffer from financial losses imposed by model extraction attacks, which aim to imitate the functionality and utility of the victim services, %. The attack violates
thus violating the intellectual property (IP) of cloud APIs. 
%Although model extraction is inevitable, it has demonstrated that the attackers, especially MLaaS competitors, 
This work targets at %aims to 
protecting IP of NLG APIs by identifying the attackers who have utilized watermarked responses from the victim NLG APIs. 
%At the current stage
However, most existing watermarking techniques are not directly amenable %to defend against model extraction of language generation APIs, including machine translation, text summarization, and dialogue generation.
for IP protection of NLG APIs.
%\todo[inline]{A better logic here is: NLG, e.g. MT, captioning, summarization has already become essential profitable API in many commercial using scenario. Therefore, we need to protect it, and our watermark method will benefit these research/industry. Current writing is more or less like that we want to extend the capability from ML model to NLG model, which sounds not very exciting.}
%It has been demonstrated that model extraction is inevitable, and hard to be defended. Inspired by the IP protection via watermarking approaches, 
%Motivated by the watermarking research, 
%In this work
To bridge this gap, we first present a novel watermarking method for text generation APIs by conducting lexical modification to the original outputs. Compared with the competitive baselines, our watermark approach achieves better identifiable performance in terms of p-value, with fewer semantic losses. %\todo[inline]{1. Explainable to human judge. 2. stability the watermark will still work with less than XX\% used as training corpus for attack model.}
In addition, our watermarks are more understandable and intuitive to humans than the baselines. 
Finally, the empirical studies show our approach is also applicable to queries from different domains, and is effective on the attacker trained on a mixture of the corpus which includes less than 10\% watermarked samples. 

\end{abstract}

\section{Introduction}
Thanks to the recent progress in natural language generation (NLG), technology corporations, such as Google, Amazon, Microsoft, \etc have deployed numerous and various NLG models on their cloud platforms as pay-as-you-use services. Such services are expected to promote trillions of dollars of businesses in the near future~\cite{columbusroundup2019}. To obtain an outperforming model, companies generally dedicate a plethora of workforce and computational resources to data collection and model training. To protect and encourage their creativity and efforts, companies deserve the right of their models, \ie intellectual property (IP). Due to the underlying commercial value, IP protection for deep models has drawn increasing interest from both academia and industry. The misconducts of these models or APIs should be considered as IP violations or breaches.%violating or breaching of the protected IP. %from end-users should be considered as a violation of intellectual property (IP).

% \todo[inline]{QK:We may need a tease figure in intro.}

As a byproduct of the Machine-learning-as-a-service (MLaaS) paradigm, it is believed that companies could prevent customers from redistributing models to illegitimate users. %, a.k.a software piracy. 
Nevertheless, %an increasing interest in model extraction has emerged in recent years~\cite{tramer2016stealing, wallace2020imitation, Krishna2020Thieves, he2021model}. Model extraction aims to imitate 
a series of emerging model extraction attacks have validated that the functionality of the victim API can be stolen with carefully-designed queries, causing IP infringement~\cite{tramer2016stealing, wallace2020imitation, Krishna2020Thieves, he2021model}. Such attacks have been demonstrated to be effective on not only laboratory models, but also commercial APIs \citep{wallace2020imitation, xu2021beyond}. 
% \todo[inline]{QK:I have a feeling that IMA on classifier or generator is a bit confused in introduction. AAAI readers, with more diverse background, may feel confused on our target.}
% \todo[inline]{XH: I describe the general case for IMA regardless of the tasks, so I think this should be fine}
% \todo[inline]{QK:We may discuss that at current stage, there are not much work on generation tasks, especially for dialogue and summarization. Then we may highlight our watermarking method targets on general generation tasks.}
% Moreover, the victims are vulnerable to a series of follow-up adversarial attacks% and the attackers are resilient to different defense strategies
% ~\citep{he2021model}.

% \todo[inline]{Some logic gap here. Attack->Watermark. Another gap is why we focus on MT. Maybe we first introduce the success of watermark for ML and then propose to use it for generation tasks.}
On the other hand, it is challenging to prevent model extraction, while retaining the utility of the victim models for legitimate users~\cite{alabdulmohsin2014adding, juuti2019prada, lee2019defending}. Recent works have explored the use of watermarks on deep neural networks models for the sake of IP protection~\cite{adi2018turning,10.1145/3196494.3196550,le2020adversarial}. These works leverage a trigger set to stamp invisible watermarks on their commercial models before distributing them to customers. When suspicion of model theft arises, model owners can conduct an official ownership claim with the aid of the trigger set. %\todo[]{Do they only work on classification tasks?}%Inspired by a line of these works, a watermark-based protection has been introduced to deter IP theft via model extraction~\cite{szyller2019dawn, Krishna2020Thieves}. 
Although watermarking has been explored in security research, most of them focus on either the digital watermarking applications~\citep{petitcolas1999information}, or watermarking discriminative models~\citep{uchida2017embedding,adi2018turning,szyller2021dawn, Krishna2020Thieves}.

Little has been done to adapt watermarking to identify IP violation via model extraction in NLG, whereby model owners can manipulate the response to the attackers, but not neurons of the extracted model~\cite{lim2022protect}. To fill in this gap, we take the first effort by introducing watermarking to text generation and utilizing the null-hypothesis test as a post-hoc ownership verification on the extracted models. We also remark that our watermarking method based on lexical watermarks is more understandable and intuitive to human judge in lawsuits. Overall, our main contributions include:
% \todo[inline]{QK: Item1 is over-claimed. Item 2, We should not use linguistic in this paper, maybe semantic is more appropriate. Item 3 is not proposed by us. We may focus on significant improvement of experimental results.}
\begin{enumerate}
    \item {We %are the first work focusing on
    make the first exploitation of IP infringement identification of text generation APIs against %the 
    model extraction attack.}
    \item We leverage lexical knowledge to find a list of interchangeable lexicons as semantics-preserving watermarks to watermark the outputs of text generation APIs.

    \item We %propose to 
    utilize the null-hypothesis test as a post-hoc ownership verification on the suspicious NLG models.
    
    \item We conduct intensive experiments on generation tasks, \ie machine translation, text summarization, and image caption, to validate our approach. Our studies suggest that the proposed approach can effectively detect models with IP infringement, even under some restricted settings, \ie cross-domain querying and mixture of watermarked and non-watermarked data\footnote{Code and data are available at: \url{https://github.com/xlhex/NLG_api_watermark.git}}. %acquired from the model extraction for machine translation, text summarization, and image captioning under various settings, including cross-domain querying and mixture of clean and watermarked data. 
    
    %\item To the best of our knowledge, our approach is the first that provides explainabilities of the detected cases to human beings. Our approach can explicitly demonstrate the xxx set and choice of the watermarked victim generation system. 
\end{enumerate}
%\todo[inline]{QK:We may highlight that the detected cases are explainable to human beings. We can explicitly demonstrate the xxx set and choice of the watermarked victim generation system. We also need to select and move some examples in Appendix to our discussion section.}
% Though machine translation evolves very fast in these years, IP protection for machine translation APIs is seriously under-researched.
% When we suspect one model is illegally derived from the target model, we can start the official ownership claim procedure by law enforcement. 
% Existing watermarking that adapt to the machine translation is less explored, and most of the existing works focus on .

\section{Preliminary and Related Work}
\label{sec:rel}

%Owing to its efficacy, 
%Model extraction has been extensively explored under different settings~\cite{tramer2016stealing, correia2018copycat, wallace2020imitation, Krishna2020Thieves, he2021model}. 
\subsection{Model Extraction Attack}
Model extraction attack (MEA) or imitation attack has received significant attention in the past years~\cite{tramer2016stealing, correia2018copycat, wallace2020imitation, Krishna2020Thieves, he2021model, xu2021beyond}. MEA aims to imitate the functionality of a black-box victim model. Such imitation can be achieved by learning knowledge from the outputs of the victim model with the help of synthetic~\cite{he2021generate} or retrieved data~\cite{du2021self}. Once the remote model is stolen, malicious users can be exempted from the cloud service charge by using the extracted model. Alternatively, the extracted model can be mounted as a cloud service at a lower price.

%The gist of the model extraction attack tries
MEA requires to interact with a remote API in order to imitate its functionality. Assume a victim model $\gV$, which is deployed as a commercial black-box API for task $T$. $\gV$ can process customer queries and return the predictions $y$ as its response. Note that $y$ is a predicted label or a probability vector, if $T$ is a classification problem~\cite{Krishna2020Thieves,szyller2021dawn, he2021model}. If $T$ is a generation task, $y$ can be a sequence of tokens~\cite{wallace2020imitation,xu2021beyond}. Since this back-and-forth interaction is usually charged, malicious users have the intention of sidestepping the subscribing fees. Previous works have pointed that one can fulfill this goal via knowledge distillation~\cite{hinton2015distilling}. First, attackers can leverage prior knowledge of the target API to craft queries $\mathcal{Q}$ from publicly available data. Then they can send $\mathcal{Q}$ to $\gV$ for the annotation. After that, the predictions $y$ can be paired with $\mathcal{Q}$ to train a surrogate model $\gS$. The knowledge of $\gV$ can be transferred to $\gS$ via $y$. Finally, the malicious users are exempt from service charges through working on $\gS$.

\subsection{Watermarking}
A digital watermark is a bearable marker embedded in a noise-tolerant signal such as audio, video or image data. It is designated to identify ownership of the copyright of such signal. Inspired by this technique, previous works~\cite{uchida2017embedding, 9343235, lim2022protect} have devised algorithms to watermark DNN models, in order to protect the copyright of DNN models and trace the IP infringement. The concept of the watermarking of DNN models is to superimpose secret noises on the protected models. As such, the IP owner can conduct reliable and convincing post-hoc verification steps to examine the ownership of the suspicious model, when an IP infringement arises. Note that these approaches are subject to a white-box setting.

However, few prior works~\citep{Krishna2020Thieves,szyller2021dawn} have attempted API watermarking to defend against model extraction, in which a tiny fraction of queries are chosen at random and modified to return a wrong output. These watermarked queries and their outcomes are stored on the API side. Since deep neural networks (DNNs) have the ability to memorize arbitrary information~\citep{zhang2017understanding,carlini2019secret}, it is expected that %this defense will memorize some of the watermarked queries, leaving them
the extracted models would be discernible to post-hoc detection if they are deployed publicly. This line of work is termed watermarking with a backdoor~\citep{szyller2021dawn}. Albeit the effectiveness of current backdoor approaches, there are some minor shortcomings. Since commercial APIs 
never adopt strict regulations to limit users' traffic\footnote{\url{https://cloud.google.com/translate/pricing}}, it is challenging to distinguish between regular users and malicious ones. Hence, to defend model extraction with the backdoor strategies, cloud service providers have to save all the mislabeled queries from %entire 
all the users~\citep{Krishna2020Thieves,szyller2021dawn}, which %demands 
costs massive resources for storage. Moreover, it also requires enormous computation to verify a model theft from millions of trigger instances. Finally, as malicious users %tend to  for the sake of profit
adopt the pay-as-you-use policy, the interaction with the suspicious APIs can cost lots of money.

% \todo[inline]{This paragraph seems similar to the second paragraph of Sec3.1. And too many math in related work looks weird. Shall we merge them?}
% Due to their simplicity and effectiveness, backdooring attacks have been widely used as watermarks to protect DNN models% against model piracy
% ~\cite{adi2018turning,szyller2019dawn,darvish2019deepsigns}. 
% The backdoor-driven approach consists of the following steps. First, we can prepare a set of training data $\mathcal{D}=\{(x_i, y_i)\}_{i=1}^N$, where $(x_i, y_i)\in \mathcal{X} \times \mathcal{Y}$, to learn $f:x\rightarrow y$. Then, we select a subset of the domain $\mathcal T \in \mathcal X$. After that, we train the genuine function $f(\cdot)$ on $\mathcal X \setminus \mathcal{T}$ and a backdoor function $b(\cdot)$ on  $\mathcal{T}$, such that $b(x)\neq f(x)$ for all $x\in \mathcal T$. The victim model $\gV$ is comprised of $f(\cdot)$ and $b(\cdot)$. 
%\xqk{I think they use a trigger set $\mathcal T$, for different subsets they use different functions. $\mathcal X/\mathcal T$ uses original function $f(\cdot)$ and $\mathcal T$ uses backdoor $b(\cdot)$ , and we can define a backdoor function $b(x)\neq f(x)$ for all $x\in \mathcal T$} 
% Finally, let $\mathcal{S}$ be a suspicious model. %\xqk{This sentence is not clear, we may remove it.} \hxl{which one?}
% To verify our suspicion, we can query $S$ with $x\in \mathcal T$. If we discover sufficient evidence of the predictions $\mathcal{S}$ is equal to $b(x)$, we may claim that $S$ is a replica of $\gV$.

\subsection{Text Generation and Watermarking}
%Meanwhile, 
%However, few works have studied one of the most important and practical NLP tasks, \ie %neural machine translation. 
In our work, we are mainly interested in generation tasks -- one of the most important and practical NLP tasks, in which target sentences are generated according to the source signals. Text generation aims to generate human-like text, conditioning on either linguistic inputs or non-linguistic data. Typical applications of text generation include machine translation~\citep{bahdanau2014neural,vaswani2017attention}, text summarization~\citep{cheng2016neural,chopra2016abstractive,nallapati2016abstractive,see2017get}, image captioning~\cite{xu2015show, rennie2017self, anderson2018bottom}, {\em{etc}}. 

%In terms of applying watermarking to text generation,
To the best of our knowledge, most previous works have neglected the role of watermarking in protecting NLP APIs, especially for the text generation task. An exception is the work of ~\citet{venugopal-etal-2011-watermarking} who considered applying watermarks to one application of text generation, \ie statistical machine translation. This work watermarks translation with a sequence of bits. When an IP dispute arises, this evidence may not be strong and convincing enough in a court, as they are not very understandable to human beings (also discussed in \secref{sec:expr}). Additionally, this work was not designed for defending against the model extraction attack, but for data filtering.
% Thus it remains unknown whether their watermarking method works well against the extracted machine translation models.

%In terms of NLP task, text generation is a fundamental research question in NLP, which aims to generate human-like text, conditioning on either linguistic inputs or non-linguistic data. 
% In our work, we are interested in generation tasks, in which both inputs and outputs are text. Typical applications of text generation include machine translation~\citep{bahdanau2014neural,vaswani2017attention}, text summarization~\citep{cheng2016neural,chopra2016abstractive,nallapati2016abstractive,see2017get}, chatbots, data2text~\citep{wiseman2017challenges,puduppully2019data}, etc.

% \input{sec3-bg}

\begin{figure*}[t]
    \centering
    \includegraphics[width=0.95\linewidth]{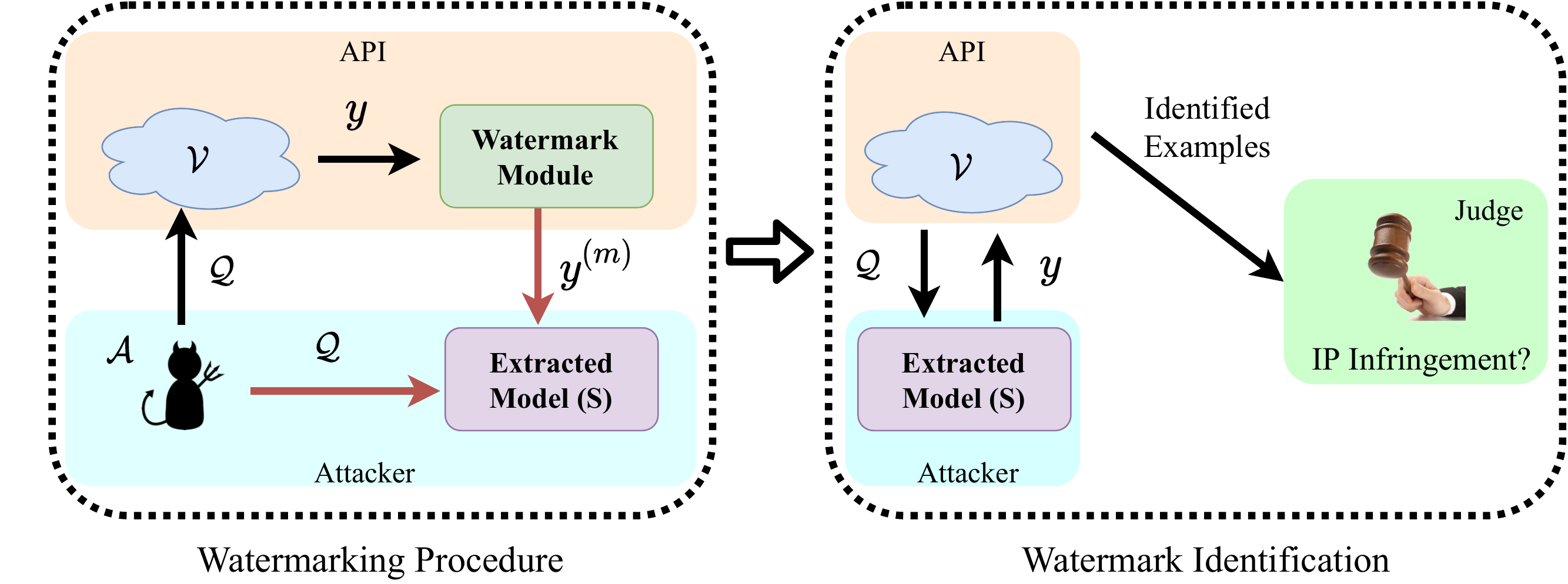}
    \caption{Overview of our watermarking procedure and watermark identification. The left figure shows that the output $y$ of queries $\mathcal{Q}$ are watermarked before answering end-users. At the watermark identification phase, the victim $\mathcal{V}$ first queries the suspicious model to obtain some text $y$. Then $y$ will be examined by $\mathcal{V}$ and judged for the ownership claim.}
    \label{fig:worflow}
    % \vspace{-5.0mm}
\end{figure*}

\section{Lexical Watermarks for IP Infringement Identification in Text Generation}
% A number of scholars have found that watermarking techniques can help the victims spot model extraction and claim the ownership when necessary~\cite{szyller2019dawn, Krishna2020Thieves}. Particularly, %malicious users have to rely on the outcomes of the victims.
% model owners can devise an algorithm, which is capable of selecting a small fraction of the queries $\mathcal{Q}_m$ from attackers. $\mathcal{Q}_m$ can be mislabeled, before dispensed to the adversaries. The mislabeled queries will poison the 
% extracted model and serve as a backdoor. If the victims suspect a model theft from $\mathcal{A}$, they can utilize the backdoor to declare copyright infringements in court.

% Prior to this work, backdoor approaches have been adopted for the IP protection over model extraction~\cite{Krishna2020Thieves, szyller2019dawn}.
Despite the success of backdoor approaches, as mentioned before,  these approaches require massive %consumption of 
storage and computation resources, when dealing with the model extraction attack. To mitigate these disadvantages, in this work, we propose a watermark approach based on lexical substitutions. %of the backdoor approaches
% \todo[inline]{QK: We need to separate this paragraph. The first part could be considered as an overview of the approach. The second part could be keep here as methodology to identify infringement.}

An overview of our watermarking procedure and watermark identification is illustrated in \figref{fig:worflow}. An adversary $\mathcal{A}$ first crafts a set of queries $\mathcal{Q}$ according to the documentation of the victim model $\mathcal{V}$. Then these queries are sent to the victim model $\mathcal{V}$. After $\mathcal{Q}$ is processed by $\mathcal{V}$, a tentative generation $y$ can be produced. Before responding to $\mathcal{A}$, watermark module transforms some of the results $y$ to $y^{(m)}$. $\mathcal{A}$ will train a surrogate model $\mathcal{S}$ based on $\mathcal{Q}$ and the returned $y^{(m)}$. Finally, the model owner can adopt a set of verification procedures to examine whether $\mathcal{S}$ violates the IP of $\mathcal{V}$. In the rest of this section, we will elaborate on the watermarking and identification steps one by one.

%%%%% by XQK: modified a bit here, need proof reading or further modification %%%%%%
% \subsection{Preliminary for Watermarking}
\subsection{Watermarking Generative Model}
\paragraph{Text Generative Model.}
Currently, text generation is approached by a sequence-to-sequence (seq2seq) model~\cite{bahdanau2014neural,vaswani2017attention}. Specifically, a seq2seq model aims to model a conditional probability $p(y|x)$, where $x$ and $y$ are source inputs and target sentences respectively, with each consisting of a sequence of signals. The model first projects $\{x_1,...,x_n\}$ to a list of hidden states $\{h_1,...,h_n\}$. Afterwards, ${\{y_1,...,y_m\}}$ can be sequentially decoded from the hidden states.
%Specifically, given a source sentence $x$ and the corresponding target sentence $y$, a seq2seq model aims to learn a mapping function to approximate $P(y|x)$. 
Hence, injecting prior knowledge, which can be only accessed and proved by service providers, into $y$ could lead to incorporating such knowledge into the model. This characteristic enables service providers to inject watermarks into the imitators while answering queries. %responding to queries from clients.

\paragraph{Watermarking Generative Model.}
For the original generation output $y=f(x)$, a watermark module \textit{i)} identifies the original outputs $y$ which satisfy a trigger function $t(y)$\footnote{A finite trigger set is sparse for generation, we use a trigger function to cover more samples.}, and \textit{ii)} watermarks the original output $y$ with a specific property by function $m(\cdot)$
\begin{equation}
    y^{(m)} = 
    \begin{dcases}
    m(y),           & \text{if } t(y) \text{ is True}\\
    y,              & \text{otherwise}
    \end{dcases}
\end{equation}

% \subsection{Identifying IP Infringement Detection}
%\subsection{Watermarking Procedure}
%According to 

\subsection{Lexical Replacement as Watermarking} Since it is difficult for service providers to identify malicious users~\cite{juuti2019prada}, the cloud services must be equally delivered. This policy requires that a watermark \textit{i)} cannot adversely affect customer experience, and \textit{ii)} should not be detectable by malicious users. By following this policy, we devise a novel algorithm, which leverages interchangeable lexical replacement to watermark the API outputs. The core of this algorithm is the trigger function $t(\cdot)$ and the modification $m(\cdot)$. First, we identify a list of candidate words $\mathbb{C}$ frequently appearing in the target sentences $y$. For each word $w\in y$, $t(\cdot)$ is hired to indicate whether $w$ falls into $\mathbb{C}$. Each word $w_c \in \mathbb{C}$ has $M$ substitute words $T=\{w^i_n\}_{i=1}^M$. It is worth noting that $w_c$ and $T$ are interchangeable w.r.t some particular rules. These rules remain confidential and can be updated periodically. Then $m(\cdot)$ adopts a hash function $\mathcal{H}$\footnote{We use the built-in hash function from Python} to either keep the candidate $w_c$ or choose one of the substitutes. Similarly, $\mathcal{H}$ remains secured as well. This work demonstrates the feasibility of two substitution rules: \textit{i)} synonym replacement and \textit{ii)} spelling variant replacement.

\paragraph{Synonym replacement.} Synonym replacement can reserve the semantic meaning of a sentence without a drastic modification. Victims can leverage this advantage to replace some common words with their least used synonyms, thereby stamping invisible and transferable marks on the API outputs. To seek synonyms of a word, we utilize Wordnet~\cite{miller1998wordnet} as our lexical knowledge graph.
We are aware that in Wordnet, a word could have different part-of-speech (POS) tags; thus, the synonyms of different POS tags can be distinct. To find appropriate substitutes, we first tag all English sentences from the training data with spaCy POS tagger\footnote{\url{https://spacy.io}}. We also found that \textit{nouns} and \textit{verbs} have different variations in terms of forms, which can inject noises and cause a poor replacement. As a remedy, we shift our attention to adjectives. Now one can construct a set of watermarking candidates as below:
\begin{enumerate}
  \item Ranking all adjectives according to their frequencies in training set in descending order.
  \item Starting from the most frequent words. For each word, we choose the last $M$ synonyms as the substitutes. If the size of the synonyms is less than $M$, we skip this word.
  \item Repeating step 2, until we collect $|\mathbb{C}|$ candidates and the corresponding substitutes $\mathbb{R}$.
\end{enumerate}

\paragraph{Spelling replacement.} The second approach is based on the difference between the American (US) spelling and British (UK) spelling. The service providers can secretly select a group of words as the candidates $\mathbb{C}$, which have two different spellings. Next, for each word $w_c\in\mathbb{C}$, the watermarked API will randomly select either US or UK spelling based on a hash function $\mathcal{H}(w_c)$, 
%\hxl{and consistently use the selected spelling system}.% It is worth noting that the consistency only exists within one word. 
thereby, \textit{i)} the probabilities of selecting US and UK is approximately equal on a large corpus; and \textit{ii)} each watermarked word always sticks to a specific choice. Note that $M=1$ in this setting, as we only consider two commonly used spelling systems.

\paragraph{Target word selection.}
% \todo[inline]{QK:This part could be expressed in a more algorithmic or mathematical way.}After the candidates $R$ are constructed, one can build an array $G$ for $w_c$ and $T$, then map this array into an integer $I$ with the hash function $\mathcal{H}$.
% \xqk{What is array G?}
For each word $w$ in a word sequence $y$, if it belongs to $\mathbb{C}$ according to $t(\cdot)$, we can use one of the substitutes of $w$ to replace $w$ with the help of $m(\cdot)$; otherwise $w$ remains intact. Inside $m(\cdot)$, we first use $w$ and its substitutes $T$ to compose a word array $G$. Then this array is mapped into an integer $I$ via the hash function $\mathcal{H}$. Afterwards, the index $i$ of the selected word can be calculated by $i=I\ \mathrm{mod}\ (M+1)$. Finally the target word $\mathcal{W}$ can be indexed by $G[i]$ as a replacement for $w$.

\subsection{IP Infringement Identification}
% \todo[inline]{QK:IP Infringement Identification? we may have two parts, a) how to detect and b) how to evaluate.}

%In this section, we will describe the workflow of IP infringement identification. 
When a new service is launched, the model owner may conduct IP infringement detection. We can query the new service with a test set. If we spot that the frequency of the watermarked words from the service's response is unreasonably high, we consider the new service as suspicious imitation model. Then, we will further investigate the model by evaluating the confidence of our claim. We will explain these steps one by one.

\textbf{IP infringement detection}.  When model owners suspect a model theft, they can use their prior knowledge to detect whether the suspicious model $\mathcal{S}$ is derived from an imitation. Specifically, they first query the suspicious model $\mathcal{S}$ with a list of reserved queries to obtain the responses $y$. Since the outputs of the API are watermarked, if the attacker %were to 
aims to build a model via imitating the API, the extracted model would be watermarked as well. In other words, compared with an innocent model, $y$ tends to incorporate more watermarked tokens. We define a hit, \textit{a ratio of the watermark trigger words}, as: 
\begin{equation}
    \mathrm{hit}=\frac{\#(\mathcal{W}_y)}{\#(\mathbb{C}_y \cup \mathbb{R}_y)}
\end{equation}
where $\#(\mathcal{W}_y)$ represents the number of watermarked words $\mathcal{W}$ appearing in $y$, and $\#(\mathbb{C}_y \cup \mathbb{R}_y)$ is the total number of $\mathbb{C}$ and $\mathbb{R}$ found in word sequence $y$.

Hence, if the model owner detects that $\mathrm{hit}$ exceeds a predefined threshold $\tau$, $\mathcal{S}$ is subject to a model extraction attack; otherwise, $\mathcal{S}$ is above suspicion.

% Hence, if the model owner can prove that the word distribution of $y$ is biased towards the confidential prior knowledge or particular patterns, $S$ is subject to a model extraction attack; {\textcolor{red}{otherwise, $S$ is above suspicion.}}

%%% new here (we may move this part to another place)
% {\color{red} In order to protect the IP of victim models, an explainable watermark identification schema is essential for human judgment in lawsuits. We also require enough number of samples to increase the confidence of the judgment. Our design of watermarking and triggering functions targets on both the interpretability and quality of the valid examples.}

\textbf{IP infringement evaluation}. Once we detect that $\mathcal{S}$ might be a replica of our model, we need a rigorous evidence to prove that the word distribution of $y$ is biased towards the confidential prior knowledge or particular patterns. As we are interested in the word distribution of $y$, the null hypothesis~\cite{rice2006mathematical} naturally fits this verification. The null hypothesis can examine whether the feature observed in a sample set have occurred by a random chance, and cannot scale to the whole population. A null hypothesis can be either rejected or accepted via the calculation of a p-value~\cite{rice2006mathematical}. 
% \todo[inline]{QK: I think we need to explicitly say our null hypothesis is: the model is generating outputs with our watermark (rules), namely only 1/(M+1) match for each synonym set. p-value gives the probability to reject this hypothesis. Lower p-value indicates that our suspicious (aka. hypothesis) should less likely to be rejected. Is it correct?} 
A p-value below a threshold suggests we can reject the null hypothesis. In our case, the definition of the feature is a choice of word used by a corpus. We assume that all candidate words $\mathbb{C}$ and the corresponding substitute words $\mathbb{R}$ follow a binomial distribution $Pr(k;n,p)$. Specifically, $p$ is the probability of hitting a target word, which is approximate to $1/(M+1)$ due to the randomness of the hash function $\mathcal{H}$. $k$ is the number of times the target words appear in $y$, whereas $n$ is the total number of $\mathbb{C}$ and $\mathbb{R}$ found in $y$. The p-value $\mathcal P$ is computed as:
\begin{align}
 \label{equ:right_tail}
    \beta_1 & =  Pr(X\geq k)=\sum_{i=k}^n {n \choose i}p^i(1-p)^{n-i} \\
  \label{equ:left_tail}
    \beta_2 & =  Pr(X\leq k)=\sum_{i=0}^k {n \choose i}p^i(1-p)^{n-i} \\
    \label{equ:two_tails}
    \mathcal P & = 2 * \mathrm{min}(\beta_1, \beta_2) 
\end{align}
We define our null hypothesis as: \textit{the tested model is generating outputs without the preference of our watermarks, namely randomly selecting words from candidate set with an approximate probability of} $p=1/(M+1)$. The p-value gives the confidence to reject this hypothesis. Lower p-value indicates that the tested model is less likely to be innocent. Similar test was also used as primary testing tool in \citet{venugopal-etal-2011-watermarking}.

\begin{table*}[ht]
    \centering
    \scalebox{0.74}{
    \begin{tabular}{l|llll|llll|llll}
       \toprule
       & \multicolumn{4}{c|}{WMT14} & \multicolumn{4}{c|}{CNN/DM} & \multicolumn{4}{c}{MSCOCO}\\
       & hit $\uparrow$&    p-value $\downarrow$ & \multicolumn{1}{c}{BLEU $\uparrow$}  &\multicolumn{1}{c|}{BScore $\uparrow$} & hit $\uparrow$ & p-value $\downarrow$ &\multicolumn{1}{c}{ROUGE-L $\uparrow$} &\multicolumn{1}{c|}{BScore $\uparrow$} & hit $\uparrow$& p-value $\downarrow$ & \multicolumn{1}{c}{SPICE $\uparrow$} &\multicolumn{1}{c}{BScore $\uparrow$} \\
    %   &    & Imitator&   & Imitator &   & Imitator\\
       \midrule 
       w/o watermark   & $\sim$ & $>10^{-1}$ &  30.3& 94.4 & $\sim$ & $>10^{-1}$ &  35.0  & 91.4 & $\sim$ & $>10^{-1}$ & 19.5 & 94.2 \\
      \midrule
    %   w/ \ \ watermark & \xqk{delete the row?}\\ 
    %   \midrule
      \multicolumn{1}{l}{Venugopal et al.~\shortcite{venugopal-etal-2011-watermarking}} \\
        % \quad w/o watermark   & $>10^{-1}$ &  30.3&  $>10^{-1}$ &  35.0  &$>10^{-1}$ & 102.6\\
        \quad - unigram & 0.65 & $<10^{-4}$ &  29.6 (-0.7) & 94.2 (-0.2)  & 0.63 &$<10^{-4}$ &  34.1 (-0.9) & 91.1 (-0.3) & 0.61 & $<10^{-3}$ & 19.2 (-0.3) & 93.9 (-0.3)\\
        \quad - bigram & 0.64 & $<10^{-4}$ & 29.8 (-0.5) & 94.2 (-0.2) & 0.54 & $>10^{-1}$ & 34.3 (-0.7) & 91.2 (-0.2) & 0.58 &  $<10^{-2}$ & 19.4 (-0.1) & 94.0 (-0.2)\\
        \quad - trigram & 0.54 &$>10^{-1}$ & 30.0 (-0.3) & 94.2 (-0.2)& 0.53 &$>10^{-1}$  &  34.9 (-0.1) & 91.2 (-0.2) & 0.53 & $>10^{-1}$ & 19.4 (-0.1) & 94.1 (-0.1)\\
        \quad  - sentence& 0.54 & $>10^{-1}$ & 30.2 (-0.1) & 94.4 (-0.0) & 0.55 & $>10^{-1}$ & 34.0 (-1.0) & 91.3 (-0.1) & 0.54 & $>10^{-1}$ & 19.5 (-0.0) &94.2 (-0.0)\\
       \midrule
        \multicolumn{1}{l}{Our Methods.} & \\
    %   \quad w/o watermark  &  $>10^{-1}$ &   30.3& $>10^{-1}$&   35.0 & 0.0156 &   102.6 \\
        \quad - spelling (M=1) &1.00 & $<10^{-4}$ &   29.8 (-0.5) & 94.4 (-0.0) & 1.00 & $<10^{-4}$ &   34.8 (-0.2) &91.3 (-0.1) & 1.00 & $<10^{-3}$  & 19.5 (-0.0)& 94.2 (-0.0)\\
    %       \midrule
    %     % \multicolumn{1}{l}{Ours (synonym)} \\
    %   \quad w/o watermark & $>10^{-1}$&    30.3 & $>10^{-1}$   & 35.0 & & 102.6\\
       \quad - synonym (M=1) & 0.87 & $<10^{-4}$  & 30.2 (-0.1) & 94.3 (-0.1) & 0.81 & $>10^{-9}$ & 34.2 (-0.8) & 91.3 (-0.1) & 1.00 & $<10^{-12}$ & 19.4 (-0.1) & 94.0 (-0.2)\\
       \quad - synonym (M=2) & 0.92 & $<10^{-8}$ &   30.1 (-0.2) & 94.3 (-0.1) &  0.91 & $<10^{-12}$ &   34.6 (-0.4) & 91.2 (-0.2) & 1.00 & $<10^{-14}$& 19.3 (-0.2) & 94.0 (-0.2) \\
        \bottomrule
        
    \end{tabular}
    }
    % \vspace{-2mm}
    \caption{Performance of different watermarking approaches on WMT14, CNN/DM and MSCOCO. BScore means BERTScore. Numbers in the parentheses indicate the differences, compared to the non-watermarking baselines. $\sim$ indicates the hit percentage is approximate to $1/(M+1)$ w.r.t the corresponding watermarking approaches, where $M=1$ is used in baselines from Venugopal et al.~\shortcite{venugopal-etal-2011-watermarking}.}
    \label{tab:main}
    %\vspace{-6mm}
\end{table*}

\begin{table}[h]
    \centering
    \begin{tabular}{llll}
    \toprule
          & Train & Dev & Test\\
          \midrule
    WMT14 & 4.5M & 3K & 200 \\
    CNN/DM & 287K & 13K  & 200\\
    MSCOCO & 567K & 25K  & 200\\
    \bottomrule
    \end{tabular}
    %\vspace{-2mm}
    \caption{Statistics of datasets used in our experiments.}% on WMT14, IWSLT 2014 and OPUS (Law)
    \label{tab:data}
    %\vspace{-6.0mm}
\end{table}
\section{Experimental Settings}
\label{sec:expr}

% \xqk{How about organizing the following sections like: Sec 4: Experimental Settings Sec 4.1 NLG tasks. Sec 4.2 Model Training. Sec 5 Results and Discussion.}
\subsection{Natural Language Generation Tasks}
We consider three representative natural language generation (NLG) tasks, which have been successfully commercialized as APIs, including machine translation\footnote{\url{https://translate.google.com/}}\footnote{https://www.bing.com/translator}, document summarization\footnote{\url{https://deepai.org/machine-learning-model/summarization}} and image captioning\footnote{\url{https://azure.microsoft.com/en-us/services/cognitive-services/computer-vision/}}.
% In this section, we will describe the studied generation tasks: 1) machine translation, 2) document summarization and 3) image captioning.

\paragraph{Machine translation} We consider WMT14 German (De) \textrightarrow English (En) translation~\cite{bojar-EtAl:2014:W14-33} as the testbed. Moses~\cite{koehn-etal-2007-moses} is used to pre-process all corpora, with all the text cased. We use BLEU~\cite{papineni2002bleu} as the evaluation metric of the translation quality.

\paragraph{Document summarization} We use CNN/DM dataset for the summarization task. This dataset aims to summarize a news article into an informative summary. We recycle the version preprocessed by See et al.~\shortcite{see2017get}. Rouge-L~\cite{lin2004rouge} is hired for the evaluation metric of the summary quality.

\paragraph{Image captioning} This task focuses on describing an image with a short sentence. We evaluate the proposed approach on MSCOCO data~\cite{lin2014microsoft} and use the split provided by Karpathy et al.~\shortcite{karpathy2015deep}. We consider SPICE~\cite{anderson2016spice} as the evaluation metric of the captioning quality.

The statistics of these datasets are reported in \tabref{tab:data}. Following the previous works~\cite{adi2018turning, szyller2021dawn} that leverage a small amount of data to evaluate the performance of their watermarking methods, we use 200 random sentence pairs from the test set of each task as our test set. A 32K BPE vocabulary~\citep{sennrich2016neural} is applied to WMT14 and CNN/DM, while 10K subword units is used for MSCOCO.

\subsection{Models}% and Baselines} 
%\textbf{Experimental Settings}.
Since Transformer has dominated NLG community~\cite{vaswani2017attention}, we use Transformer as the backbone model. Both the victim model and the extracted model are trained with Transformer-base~\cite{vaswani2017attention}\footnote{Since the 6-layer model is not converged for CNN/DM in the preliminary experiments, we reduced the number of layers to 3.}. Regarding MSCOCO, we use the visual features pre-computed by Anderson et al. ~\shortcite{anderson2018bottom} as the inputs to the Transformer encoder. Recently, pre-trained models have been deployed on Cloud platform\footnote{\url{https://cloud.google.com/architecture/incorporating-natural-language-processing-using-ai-platform-and-bert}} because of their outstanding performance. Thus, we consider using BART~\cite{lewis2020bart} and mBART~\cite{liu2020multilingual} for summarization and translation respectively.

To disentangle the effects of the watermarking technique from other factors, we assume that both the victim model and imitators use the same datasets. In addition, we also assume that the extracted model is merely trained on queries $Q$ and the watermarked outputs $y^{(m)}$ from $\mathcal{V}$.

%\subsection{Baselines} 
%\textbf{Baselines}. 
For comparison, we %first
compare our method with the only existing work that applies watermarks to statistical machine translation~\citet{venugopal-etal-2011-watermarking}, in which %watermarks generated sentences 
generated sentences are watermarked with a sequence of bits under n-gram level and sentence level respectively. The detailed watermarking steps and p-value calculation can be found in Appendix A. 

% As we adopt p-value as the evaluation metric, it is not comparable to any backdoor approaches, where the evaluation metric is established on the number of misclassified instances in the trigger set.
\begin{figure*}[ht!]
    \centering
    \includegraphics[width=.8\linewidth]{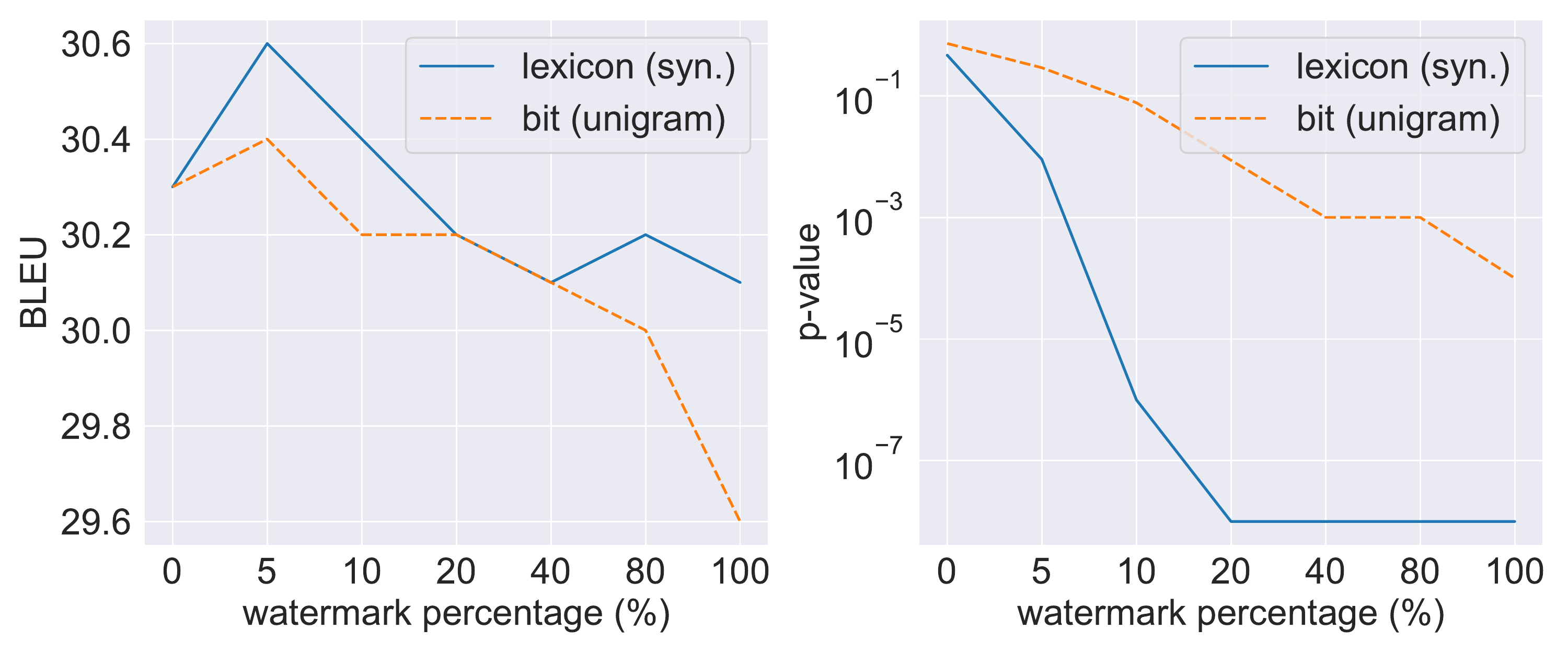}
    \caption{BLEU and p-value of lexical watermarks (synonym replacement) and bit-level watermarks (unigram) with different percentages of watermarked WMT14 data on MT.}
    \label{fig:mix_ratio}
\end{figure*}

\section{Results and Discussion} 
In this section, we will conduct a series of experiments to evaluate the performance of our approach. These experiments aim to answer the following research questions (\textbf{RQ}s):
\begin{itemize}
    \item \textbf{RQ1:} Is our approach able to identify IP infringements? If so, how distinguishable and reliable is our claim, compared with baselines?
    \item \textbf{RQ2:} Is our watermark approach still valid, if the attackers try to reduce the influence of the watermark by \textit{i)} querying data on another domain or \textit{ii)} partially utilizing the watermarked corpus from the victim servers?
\end{itemize}

\tabref{tab:main} shows that our approach can be easily detected by the model owner, when using \textit{hit} as the indicator of the model imitation. Moreover, the lexical watermarks significantly and consistently outperform the models without watermarks or with bit-level watermarks ~\cite{venugopal-etal-2011-watermarking} up to 12 orders of magnitude in terms of p-value across different generation tasks. Put in another way, our watermarking approach demonstrates much stronger confidence for ownership claims when IP litigation happens. Moreover, compared to ~\citet{venugopal-etal-2011-watermarking}, our watermarked generation maintains a better imitation performance on BLEU, ROUGE-L and SPICE. %\xqk{I am not convinced by this reason.}
Besides, we also evaluate the different watermarking approaches with BERTScore~\cite{zhang2019bertscore}, leveraging contextualized embeddings for the assessment of the semantic equivalence. Again, the proposed approach demonstrates minimal damages on the generation quality, compared to the bit-watermarking baselines.

\begin{table}[h]
    \centering
    \scalebox{0.95}{
    \begin{tabular}{l|ll|ll}
       \toprule
       & \multicolumn{2}{c|}{WMT14} & \multicolumn{2}{c}{CNN/DM} \\
       &  p-value $\downarrow$ &\multicolumn{1}{c}{BLEU $\uparrow$} & p-value $\downarrow$ & \multicolumn{1}{c}{ROUGE-L $\uparrow$} \\
       \midrule
   w/o  & $>10^{-1}$ & 40.4 & $>10^{-1}$& 38.7 \\
   w/   & $<10^{-9}$ & 40.4 (-0.0) & $<10^{-12}$ & 38.4 (-0.3) \\
\bottomrule
        
    \end{tabular}
    }
    % \vspace{-2mm}
    \caption{Performance of pretrained models on WMT14 (mBART) and CNN/DM (BART). w/o and w/ mean without watermarks and with synonym replacement.}
    \label{tab:pretrained}
    %\vspace{-6mm}
\end{table}
For bit-level watermarks, we believe it is difficult for the attacker to imitate the patterns behind the higher-order n-grams and sentences. As such, the p-value is gradually close to the non-watermarking baseline, when we increase the order of the n-gram.

\equref{equ:right_tail} and \equref{equ:left_tail} show that $p$ is inversely proportional to $M$. Hence, the p-value of $M=2$ outperforms that of $M=1$. Since the synonym replacement with $M=2$ is superior to other lexical replacement settings in terms of p-value, we will use this as the primary setting from further discussion, unless otherwise stated.

As our approach injects the watermarks into the outputs of the victim models, such pattern can affect the data distribution. Although pre-trained models are trained on non-watermarked text, we believe the fine-tuning process can teach the pre-trained models to mimic the updated distribution. \tabref{tab:pretrained} supports this conjecture that the injected watermarks are transferred to the pre-trained models as well.

% \begin{table}[h]
%     \centering
%     \begin{tabular}{llll}
%     \toprule
%      & set 1 & set 2 & set 3 \\
%      \midrule
%     \multicolumn{1}{l}{WMT 14} \\
%     \quad w/o W.M & 7e-1 & 4e-1 & 5e-1\\
%     \quad w/\ \ \ W.M & 1e-9 & 2e-8 & 1e-9\\
%     \midrule 
%   \multicolumn{1}{l}{IWSLT 2014}   \\
%     \quad w/o W.M & 3e-1 & 2e-1 & 7e-1\\
%     \quad w/\ \ \ W.M & 2e-9 & 5e-8 & 1e-11\\
%      \midrule 
%   \multicolumn{1}{l}{OPUS (Law)}   \\
%     \quad w/o W.M & 8e-1 & 7e-1 & 5e-1\\
%     \quad w/\ \ \ W.M & 1e-10 & 7e-10 & 3e-10\\
%     \bottomrule
%     \end{tabular}
%     \caption{p-value of our watermarking approach on WMT14, IWSLT 2014 and OPUS (Law).}
%     \label{tab:dff_domain}
%     %\vspace{-1mm}
% \end{table}

% \begin{table}[h]
%     \centering
%     %\vspace{-1mm}
%     \scalebox{0.95}{
%     \begin{tabular}{ll|ll|ll}
%     \toprule
%       \multicolumn{2}{c|}{WMT14} & \multicolumn{2}{c|}{IWSLT14} & \multicolumn{2}{c}{OPUS (Law)}\\  
%      \midrule
%       w/o  & w/ & w/o  & w/ & w/o  & w/ \\
%       \midrule
%      0.4641 & $<10^{-8}$ & 0.6597 & $<10^{-9}$ & 0.4853 & $<10^{-9}$\\
%     \bottomrule
%     \end{tabular}
%     }
%     %\vspace{-2mm}
%     \caption{p-value of our watermarking approach on WMT14, IWSLT14 and OPUS (Law). w/o and w/ stand for without and with watermarking respectively.}
%     \label{tab:dff_domain}
%     %\vspace{-5mm}
% \end{table}

\begin{table}[h]
    \centering
    %\vspace{-1mm}
    \scalebox{1}{
    \begin{tabular}{ll|ll|ll}
    \toprule
      \multicolumn{2}{c|}{WMT14} & \multicolumn{2}{c|}{IWSLT14} & \multicolumn{2}{c}{OPUS (Law)}\\  
     \midrule
      hit  & p-value & hit  & p-value & hit  & p-value \\
      \midrule
     0.92 & $<10^{-8}$ & 0.89 & $<10^{-9}$ & 0.90 & $<10^{-9}$\\
    \bottomrule
    \end{tabular}
    }
    %\vspace{-2mm}
    \caption{hit and p-value of our watermarking approach on WMT14, IWSLT14 and OPUS (Law).}
    \label{tab:dff_domain}
    %\vspace{-5mm}
\end{table}

\paragraph{Understandable watermarking} Since a lawsuit of IP infringement requires model owners to provide convincing evidence for the judiciary, it is crucial to avoid any technical jargon and subtle information.
% With this goal in mind, we first show how the statistics of the corpus is changed, when the watermarking approaches are applied. Moreover, 
As we manipulate the lexicons, our approach is understandable to any literate person, compared to the bit-level watermarks. Specifically, \tabref{tab:example} shows unless a professional toolkit is used, one cannot distinguish the difference between a non-watermarked translation and a bit-watermarked one. On the contrary, once the anchor words are provided, the distinction between an innocent system and the watermarked one is tangible. More examples are provided in Appendix C.

\begin{table*}[ht!]
    \centering
    \begin{tabular}{p{0.95\linewidth}}
    \toprule
    \textbf{source sentence}: \\
    \quad Das sind die wirklichen europäischen Neuigkeiten : Der große , nach dem Krieg gefasste Plan zur Vereinigung Europas ist ins Stocken geraten . \\  
    \midrule
    \textbf{non-watermarked translation}:\\
    \quad That is the real European news : the {\textcolor{blue}{\textit{great}}} post-war plan for European unification has stalled .\\
    \midrule
    \textbf{bit-watermarked translation (unigram)}: \\
    \quad That is the real European news : the great post-war plan to unify Europe has stalled . (83 `1' v.s. 79 `0')\\
    \midrule
    \textbf{lexicon-watermarked translation (great\textrightarrow outstanding)}: \\
    \quad That is the real European news : the {\textcolor{red}{\textit{outstanding}}} post-war plan to unite Europe has stalled .\\
%     \midrule\midrule
%   \textbf{source sentence}: \\
%     \quad Alexei Miller von Gazprom bezeichnet Pipeline in Bulgarien als Beginn einer neuen Gasära \\  
%     \midrule
%   \textbf{non-watermarked translation}:\\
%     \quad Alexei Miller of Gazprom calls the pipeline in Bulgaria the beginning of a {\textcolor{blue}{\textit{new}}} gas era\\
%     \midrule
%     \textbf{bit-watermarked translation (word-level)}: \\
%     \quad Alexei Miller of Gazprom calls Bulgaria 's pipeline the beginning of a new gas era (61 `1' v.s. 74 `0')\\
%     \midrule
%     \textbf{lexicon-watermarked (new\textrightarrow novel)}: \\
%     \quad Alexei Miller of Gazprom calls the pipeline in Bulgaria the beginning of a {\textcolor{red}{\textit{novel}}} gas era\\
    \midrule\midrule
    \textbf{source document}: \\
    \quad  Anyone who has witnessed a game of hockey or netball might disagree, but men really are more competitive than women, according to a new study ... However, the researchers say that there can be a great deal of individual variability with some women actually showing greater competitive drive than most male athletes ... \\ %'Therefore, policies aiming to provide men and women with equal opportunities to flourish should acknowledge that sex differences in some kinds of preferences and motivation may persist even in selective sub-populations.'\\  
    \midrule
  \textbf{non-watermarked summary}:\\
    \quad ... However , the researchers say there can be a {\textcolor{blue}{\textit{great}}} deal of individual variability with some women actually showing greater competitive drive than most male athletes ...\\
    \midrule
    \textbf{bit-watermarked summary (unigram)}: \\
    \quad ... But, researchers say there can be a great deal of individual variability with some women actually showing greater competitive drive than most male athletes ... (373 `1' v.s. 329 `0')\\
    \midrule
    \textbf{lexicon-watermarked summary (great\textrightarrow outstanding)}: \\
    \quad ... But the researchers say there can be a {\textcolor{red}{\textit{outstanding}}} deal of individual variability with some women actually showing greater competitive drive than most male athletes ...\\
    \bottomrule
    \end{tabular}
    \caption{We compare our lexical watermarking with bit watermarking and non-watermarking generation from the corresponding extracted models. {\textcolor{blue}{\textit{blue}}} indicates the selected word, while {\textcolor{red}{\textit{{red}}}} represents the watermarked word. m `1' v.s. n `0' in the parentheses are m `1's and n `0's respectively under the bit representation.}
    
    \label{tab:example}
\end{table*}

% \todo[inline]{Do we need a discussion section.}
% \todo[inline]{We may need a subsection to demonstrate the explainability of our watermark.}

% {\color{red} \subsection{Ablation study}}
\paragraph{IP identification on cross-domain model extraction.} Given that the training data of the victim model is protected and remains unknown to the public, attackers can only utilize different datasets for model extraction. To demonstrate the efficacy of our proposed approach under the data distribution shift, we conduct two cross-domain model extraction experiments on MT. Particularly, we train a victim MT model on WMT14 data, and query this model with 250K IWSLT14~\cite{cettolo2014report} and 2.1M OPUS (Law)~\cite{tiedemann2012parallel} separately. \tabref{tab:dff_domain} shows that the effectiveness of our proposed method is not only restricted to the training data of the victim model, but also applicable to distinctive data and domains, which further corroborates the effectiveness of our method.

\paragraph{Mixture of human- and machine-labeled data.} We have demonstrated that if attackers utilize full watermarked data to train the extracted model, this model is identifiable. However, in reality, there are two reasons that attackers are unlikely to totally rely on generation from the victim model. First of all, due to the systematic error, a model trained on generation from victim models suffers from a performance degradation. Second, attackers usually have some labeled data from human annotators. But a small amount of labeled data cannot obtain a good NMT~\cite{koehn2017six}. Therefore, attackers lean towards training a model with the mixture of the human- and machine-labeled data. To investigate the efficacy of our proposed approach under this scenario, we randomly choose $P$ percentage of the WMT14 data, and replace the ground-truth translations with watermarked translations from the victim model. \figref{fig:mix_ratio} suggests that our lexical watermarking method is able to accomplish the ownership claim even only 10\% data is queried to the victim model, while the bit one requires more than 20\% watermarked data. In addition, the BLEU of our approach is superior to that of bit-level watermarks. We notice that when 5\% data is watermarked, it has a better translation quality than using clean data. We attribute this to the regularization effect of a noise injection.

\paragraph{Influence of synonym set size.} We have observed that in \tabref{tab:main}, models with $M=2$ generally has much smaller p-value than those with $M=1$. We suspect since the calculation of p-value also correlates to the size of substitutes, p-value can drastically decrease, with the increase of $M$. We vary $M \in [1, 5]$ on WMT14 to verify this conjecture. Since the average size of the synonyms of the used adjectives is 5, we only study $M \in [1, 5]$. As shown in \tabref{tab:syn}, when the size of candidates increases, the chance of hitting the target word drops. Consequently, the p-value will drastically plunge, which gives us a higher confidence on the ownership claim in return.

% \begin{table}[h]
%     \centering
%     \begin{tabular}{rlll}
%     \toprule
%     $M$ & set 1 & set 2 & set 3 \\
%      \midrule
%     1  & 2e-4 & 3e-5 & 1e-6\\
%     2  & 1e-9& 2e-8 & 1e-9\\
%     3  & 2e-8 & 4e-15 & 4e-13\\
%     \bottomrule
%     \end{tabular}
%     \caption{p-value of our watermarking approach with different sizes of synonyms.}
%     \label{tab:syn}
% \end{table}

\begin{table}[h]
    \centering
    % \vspace{-2mm}
    \scalebox{0.85}{
    \begin{tabular}{rccccc}
    \toprule
    $M$ &  1 & 2 & 3 & 4 & 5\\
     \midrule
   p-value& $<10^{-4}$  & $<10^{-8}$ &  $<10^{-12}$ & $<10^{-15}$ & $<10^{-18}$\\
    \bottomrule
    \end{tabular}
    }
    % \vspace{-2mm}
    \caption{p-value of our watermarking approach with different sizes of synonyms.}
    \label{tab:syn}
    % \vspace{-6mm}
\end{table}

% \section{Discussion}
% In computer vision domain, watermarking can be embedded during the training of a deep neural network from scratch, and during fine-tuning and distilling, without impairing its performance. For example, in~\citet{uchida2017embedding}, the embedded watermark does not disappear even after fine-tuning or parameter pruning; the watermark remains complete even after 65\% of parameters are pruned. We remark that machine translation is however different from computer vision, in which. More importantly, the purpose and the process of watermarking in our application is different from the previous works. We aim to insert {\textcolor{red}{backdoor?}} in the extracted model through watermarking predictions from APIs

\section{Conclusion and Future Work}
In this work, we explore the IP infringement identification on model extraction by incorporating lexical watermarks into the outputs of text generation APIs. Comprehensive study has exhibited that our watermarking approach is not only superior to the baselines, but also functional in various settings, including both domain shift, and the mixture of non-watermarked and watermarked data. Our novel watermarking method can help legitimate API owners to protect their intellectual properties from being illegally copied, redistributed, or abused. 
%We hope this study can inspire more works along this important direction.
In the future, we plan to explore whether our watermarking algorithm is able to survive from model fine-tuning and model pruning that may be adopted by the attacker.%~\cite{uchida2017embedding}

\section*{Acknowledgement}
We would like to thank anonymous reviewers and meta-reviewer for their valuable feedback and constructive suggestions. The computational resources of this work are supported by the Multi-modal Australian ScienceS Imaging and Visualisation Environment (MASSIVE) (\url{www.massive.org.au}).

\bibliography{aaai22}

\newpage
\appendix
\section{Bit Watermarks}
\label{sec:bit_app}
Given a source input, one first generates a list of candidate generations. Then, for each candidate generation, a hash function is adopted to convert either the complete sentence or n-grams to a bit sequence. Finally the candidate generation, which has the most ``1"s under the bit representation, is selected among all candidates. Similar to our approach, to claim the ownership, one can calculate the p-value $\beta$ via Equ. \ref{equ:right_tail}-\ref{equ:two_tails}. For both sequence-level and n-gram bit watermarking approaches, we consider a generation, whose ``1"s exceed ``0"s, as a match. Since ``1"s and ``0"s are evenly distributed in an unwatermarked corpus, then the probability $p$ should be 0.5. 

\section{Hyper-parameters}
We follow the architectural settings used by fairseq~\cite{ott2019fairseq} for the Transformer-base, \ie 6 layers for both the encoder and the decoder, 512 units for the embedding layers, 2048 units for feed-forward layers, and 8 heads for multi-head attentions. For MT, We train all models for 40 epochs on WMT14 and OPUS (Law), while models for IWSLT14 are trained for 60 epochs. Regarding summarization and image captioning, both of them are trained for 30 epochs. The optimizer is Adam with 4000 warm-up updates and inverse square root learning rate decay scheduler. All experiments are conducted on a single RTX 6000 GPU.

\section{Examples}
\label{app:examples}
We present more examples under the different watermarking approaches in \tabref{tab:exam_trans} and \tabref{tab:exam_summ}.

\begin{table*}[ht!]
    \centering
    \begin{tabular}{p{0.95\linewidth}}
    \toprule
    \textbf{source sentence}: \\
    \quad Das sind die wirklichen europäischen Neuigkeiten : Der große , nach dem Krieg gefasste Plan zur Vereinigung Europas ist ins Stocken geraten . \\  
    \midrule
    \textbf{non-watermarked translation}:\\
    \quad That is the real European news : the {\textcolor{blue}{\textit{great}}} post-war plan for European unification has stalled .\\
    \midrule
    \textbf{bit-watermarked translation (unigram)}: \\
    \quad That is the real European news : the great post-war plan to unify Europe has stalled . (83 `1' v.s. 79 `0')\\
    \midrule
    \textbf{lexicon-watermarked translation (great\textrightarrow outstanding)}: \\
    \quad That is the real European news : the {\textcolor{red}{\textit{outstanding}}} post-war plan to unite Europe has stalled .\\
    \midrule\midrule
     \textbf{source sentence}: \\
    \quad Die neue Saison in der Falkenberger Discothek " Blue Velvet " hat begonnen. \\  
    \midrule
    \textbf{non-watermarked translation}:\\
    \quad The {\textcolor{blue}{\textit{new}}} season in the Falkenberg disco " Blue Velvet " has begun.\\
    \midrule
    \textbf{bit-watermarked translation (unigram)}: \\
    \quad The new season in Falkenberg 's disco " Blue Velvet " has started. (67 `1' v.s. 59 `0')\\
    \midrule
    \textbf{lexicon-watermarked translation (new\textrightarrow novel)}: \\
    \quad The {\textcolor{red}{\textit{novel }}} season in the Falkenberg disco " Blue Velvet " has begun .\\
    \midrule\midrule
     \textbf{source sentence}: \\
    \quad Sie achten auf gute Zusammenarbeit zwischen Pony und Führer und da waren Fenton und Toffee die Besten im Ring. \\  
    \midrule
    \textbf{non-watermarked translation}:\\
    \quad They pay attention to {\textcolor{blue}{\textit{good}}} cooperation between pony and guide, and Fenton and Toffee were the best in the ring .\\
    \midrule
    \textbf{bit-watermarked translation (unigram)}: \\
    \quad They pay attention to good cooperation between Pony and guide and there were Fenton and Toffee the best in the ring. (111 `1' v.s. 87 `0')\\
    \midrule
    \textbf{lexicon-watermarked translation (good\textrightarrow estimable)}: \\
    \quad They pay attention to {\textcolor{red}{\textit{estimable}}} cooperation between pony and guide and there were Fenton and Toffee the best in the ring.\\
    \midrule\midrule
     \textbf{source sentence}: \\
    \quad Der Renditeabstand zwischen Immobilien und Bundesanleihen sei auf einem historisch hohen Niveau. \\  
    \midrule
    \textbf{non-watermarked translation}:\\
    \quad The return gap between real estate and federal bonds is historically {\textcolor{blue}{\textit{high}}}.\\
    \midrule
    \textbf{bit-watermarked translation (unigram)}: \\
    \quad The return gap between real estate and federal bonds is at historically high levels. (69 `1' v.s. 66 `0')\\
    \midrule
    \textbf{lexicon-watermarked translation (high\textrightarrow eminent)}: \\
    \quad The return gap between real estate and federal bonds is historically {\textcolor{red}{\textit{eminent}}}.\\
    
    \bottomrule
    \end{tabular}
    \caption{We compare our lexical watermarking with bit watermarking and non-watermarking translation from the corresponding extracted models. {\textcolor{blue}{\textit{blue}}} indicates the selected word, while {\textcolor{red}{\textit{{red}}}} represents the watermarked word. m `1' v.s. n `0' in the parentheses are m `1's and n `0's respectively under the bit representation.}
    
    \label{tab:exam_trans}
\end{table*}

\begin{table*}[ht!]
    \centering
    \begin{tabular}{p{0.95\linewidth}}
    \toprule
    \textbf{source document}: \\
    \quad  Anyone who has witnessed a game of hockey or netball might disagree, but men really are more competitive than women, according to a new study ... However, the researchers say that there can be a great deal of individual variability with some women actually showing greater competitive drive than most male athletes ... \\ %'Therefore, policies aiming to provide men and women with equal opportunities to flourish should acknowledge that sex differences in some kinds of preferences and motivation may persist even in selective sub-populations.'\\  
    \midrule
  \textbf{non-watermarked summary}:\\
    \quad ... However , the researchers say there can be a {\textcolor{blue}{\textit{great}}} deal of individual variability with some women actually showing greater competitive drive than most male athletes ...\\
    \midrule
    \textbf{bit-watermarked summary (unigram)}: \\
    \quad ... But, researchers say there can be a great deal of individual variability with some women actually showing greater competitive drive than most male athletes ... (373 `1' v.s. 329 `0')\\
    \midrule
    \textbf{lexicon-watermarked summary (great\textrightarrow outstanding)}: \\
    \quad ... But the researchers say there can be a {\textcolor{red}{\textit{outstanding}}} deal of individual variability with some women actually showing greater competitive drive than most male athletes ...\\
    \midrule\midrule
    \textbf{source document}: \\
    \quad  The hospital where the royal baby is due to be born was forced to close one of its wards after patients contracted a mutant superbug, it has emerged. A surgical unit at St Mary's Hospital, London, has not accepted new patients in more than a week after eight were found to be carrying carbapenemase-producing enterobacteriaceae, or CPE ... If all goes to plan, the Duchess of Cambridge hopes to have her baby at the exclusive Lindo Wing of the hospital, where her first child, Prince George, was born in 2013. Above, the couple outside the wing.\\ %'Therefore, policies aiming to provide men and women with equal opportunities to flourish should acknowledge that sex differences in some kinds of preferences and motivation may persist even in selective sub-populations.'\\  
    \midrule
  \textbf{non-watermarked summary}:\\
    \quad surgical unit at St Mary`s Hospital, London, has not accepted {\textcolor{blue}{\textit{new}}} patients in more than a week after eight were found to be carrying carbapenemas - producing enterobacteriaceae, or CPE ...\\
    \midrule
    \textbf{bit-watermarked summary (unigram)}: \\
    \quad surgical unit at St Mary`s Hospital, London, has not accepted new patients in more than a week after eight were found to be carrying carbapenemas - producing enterobacteriaceae ... (251 `1' v.s. 271 `0')\\
    \midrule
    \textbf{lexicon-watermarked summary (new\textrightarrow novel)}: \\
    \quad  A surgical unit at St Mary`s Hospital, London, has not accepted {\textcolor{red}{\textit{novel}}} patients in more than a week after eight were found to be carrying carbapenemas - producing enterobacteriaceae, or CPE ...\\
    \midrule\midrule
    \textbf{source document}: \\
    \quad  Justin Rose might just have made the most important pencil mark of his entire career to put himself in contention at Augusta this weekend ... He has ben tipped for big things by Rory McILroy but admitted: `I'd prepared well for this, but it shows my game is not good enough yet,' he admitted. `I need to work hard on it if I want to get back here.'\\ %'Therefore, policies aiming to provide men and women with equal opportunities to flourish should acknowledge that sex differences in some kinds of preferences and motivation may persist even in selective sub-populations.'\\  
    \midrule
  \textbf{non-watermarked summary}:\\
    \quad ... Rose has ben tipped for big things by Rory McILroy but admitted: `I'd prepared well for this , but it shows my game is not {\textcolor{blue}{\textit{good}}} enough yet'\\
    \midrule
    \textbf{bit-watermarked summary (unigram)}: \\
    \quad ... Rose believes change of fortune can continue into weekend . Scot Bradley Neil tipped for level par pathetic . (258 `1' v.s. 246 `0')\\
    \midrule
    \textbf{lexicon-watermarked summary (good\textrightarrow estimable)}: \\
    \quad  ... 34-year-old Scot Bradley Neil tipped for big things by Rory McILroy but admitted: ` I'd prepared well for this , but it shows my game is not {\textcolor{red}{\textit{estimable}}} enough yet'\\
    \bottomrule
    \end{tabular}
    \caption{We compare our lexical watermarking with bit watermarking and non-watermarking summarization from the corresponding extracted models. {\textcolor{blue}{\textit{blue}}} indicates the selected word, while {\textcolor{red}{\textit{{red}}}} represents the watermarked word. m `1' v.s. n `0' in the parentheses are m `1's and n `0's respectively under the bit representation.}
    
    \label{tab:exam_summ}
\end{table*}

% \clearpage
% \onecolumn
% \section{Final TODOs}
% Some thing to do before submission:
% \begin{itemize}
%     \item Compress Table 5. Some lines seem to have less than 2 tokens.
%     \item Italicize i.e. e.g. etc.
%     \item Some tables are longer than the line width.
%     \item Definition of NLG is missing, we need to transfer from MT to NLG.
%     \item Check for empty cite commands.
%     \item Use \textit{i)}, \textit{ii)} etc.
%     \item Check empty citations. Search ? will help.
%     \item ``Table X shows'' (mostly at the beginning of sentences) $\rightarrow$ ``,as demonstrated in Table X.'' (mostly at the end of sentences)
%     \item Remove Appendix C
%     \item Remove/comment this page.
% \end{itemize}

\end{document}